\documentclass[10pt,twocolumn,notitlepage,aps,pra,showpacs,superscriptaddress,floatfix]{revtex4-1}

\usepackage[colorlinks=true,citecolor=blue,urlcolor=blue]{hyperref}
\usepackage[utf8]{inputenc}
\usepackage[T1]{fontenc}
\usepackage[sc,osf]{mathpazo}
\usepackage{amsmath, amsthm, amssymb,amsfonts}
\usepackage{graphicx}
\usepackage{dcolumn}
\usepackage{bm}
\usepackage{bbm}
\usepackage{mathtools}
\usepackage{comment}
\usepackage{color}
\usepackage{natbib}

\usepackage{graphicx}
\usepackage{amssymb}
\usepackage{amsmath}
\usepackage{amsthm}
\usepackage{color}
\usepackage{psfrag}
\usepackage{epsfig}
\usepackage{bbm}
\usepackage{bm}
\hypersetup{breaklinks}

\usepackage{qcircuit}

\usepackage{soul}

\newcommand{\ket}[1]{| #1 \rangle}

\newcommand{\beq}{\begin{eqnarray}}
\newcommand{\eeq}{\end{eqnarray}}

\makeatletter
\newsavebox{\@brx}
\newcommand{\llangle}[1][]{\savebox{\@brx}{\(\m@th{#1\langle}\)}%
  \mathopen{\copy\@brx\kern-0.5\wd\@brx\usebox{\@brx}}}
\newcommand{\rrangle}[1][]{\savebox{\@brx}{\(\m@th{#1\rangle}\)}%
  \mathclose{\copy\@brx\kern-0.5\wd\@brx\usebox{\@brx}}}
\makeatother

\newcommand{\LL}{\mathcal{L}}

\newcommand{\HH}{\mathcal{H}}

\newcommand{\M}{\mathcal M}

\newtheorem*{theorem*}{Theorem}

\newtheorem{definition}{Definition}

\DeclareMathOperator{\tr}{tr}
\DeclareMathOperator{\id}{\mathbb{I}}

\begin{document}
\title{Communication through quantum-controlled noise}

\author{Philippe Allard Gu\'{e}rin}
\affiliation{Vienna Center for Quantum Science and Technology (VCQ), Faculty of Physics, University of Vienna, Boltzmanngasse 5, 1090 Vienna, Austria}
\affiliation{Institute for Quantum Optics and Quantum Information (IQOQI), Austrian Academy of Sciences, Boltzmanngasse 3, 1090 Vienna, Austria}
\author{Giulia Rubino}
\affiliation{Vienna Center for Quantum Science and Technology (VCQ), Faculty of Physics, University of Vienna, Boltzmanngasse 5, 1090 Vienna, Austria}

\author{\v{C}aslav Brukner}
\affiliation{Vienna Center for Quantum Science and Technology (VCQ), Faculty of Physics, University of Vienna, Boltzmanngasse 5, 1090 Vienna, Austria}
\affiliation{Institute for Quantum Optics and Quantum Information (IQOQI), Austrian Academy of Sciences, Boltzmanngasse 3, 1090 Vienna, Austria}

\date{\today}

\begin{abstract}
In a recent series of works~\cite{Ebler2017,Salek2018, Chiribella2018}, it has been shown that the quantum superposition of causal order -- the quantum switch --  offers an enhancement of classical and quantum channel capacity through noisy channels, a phenomenon that was coined `causal activation'. In this paper, we attempt to clarify the nature of the advantage, by comparing the quantum switch to a class of processes that can be interpreted as \textit{quantum superposition of processes with the same causal order}. We show that some of these processes can match or even outperform the quantum switch at enhancing classical and quantum channel capacity, and argue that they require the same resources as the switch. We conclude, in agreement with Abbott et al.~\cite{Abbott2018}, that the aforementioned advantages appear to be attributable to the ability to coherently control quantum operations, and not to indefinite causal order per se.


\end{abstract}
\maketitle

\section{Introduction}
Quantum mechanics, as it is usually formulated, operates on a definite causal structure. There has been significant recent interest in investigating physically relevant situations in which the causal ordering between events is itself governed by quantum mechanics~\cite{Hardy2005, Hardy2009, Chiribella2013, Oreshkov2012}. One interesting such causal structure is the quantum superposition of causal orders, also known as the \textit{quantum switch}~\cite{Chiribella2013}. In the quantum switch, a quantum system controls the order of applications of two quantum channels: when the state of the control qubit is $|0\rangle$, the channel $\M_A$ is applied on a target system before $\M_B$, while if the control qubit is $|1\rangle$, $\M_B$ is applied before $\M_A$. The control qubit can be the path of a photon in an interferometer, but more exotic realisations where gravity plays a role have also been studied~\cite{Zych2017, Dimic2017, Guerin2018}. It has been established theoretically that the quantum switch offers various information processing advantages over causally ordered quantum mechanics~\cite{Chiribella2012, Araujo2014_PRL, Guerin2016}, and there have been multiple experimental implementations of the switch~\cite{Procopio2015, Rubino2017, Rubino2017ent, Goswami2018_switch, Goswami2018_comm, Wei2018, Guo2018}

More recently, a series of papers~\cite{Ebler2017, Salek2018,Chiribella2018} have investigated the role of the switch in the context of classical and quantum Shannon information theory, and they have shown that the quantum switch offers advantages for enhancing channel capacity through noisy channels. All three tasks in Refs.~\cite{Ebler2017, Salek2018,Chiribella2018} have the same basic structure:
\begin{enumerate}
\item Two \textit{fixed} noisy channels $\M_A, \M_B$ are chosen.
\item A process -- a bilinear supermap -- turns these into a quantum channel $\mathcal{W}(\M_A, \M_B)$. The specific process that is considered is the quantum switch.
\item The information transmitting properties of $\mathcal{W}$ are studied and it is concluded on these grounds that, for example, that `\textit{when the order of the communication channels in between them is a quantum degree of freedom, it can increase the entanglement and enable transmission of quantum information beyond what was possible in a definite causal order.'}~\cite{Salek2018}
\end{enumerate}
The effects described in Ref.~\cite{Ebler2017, Salek2018} have been reproduced experimentally~\cite{Goswami2018_comm, Guo2018}.

In Refs.~\cite{Ebler2017, Salek2018}, the authors are comparing the switch against a class of processes in which $\M_A$ is applied, say, before $\M_B$, and there is no available `side channel' that could be used to transmit more information; in this paper we will call them \textit{direct processes}. In Ref.~\cite{Chiribella2018}, the authors also consider `independent channels in a superposition of alternative paths'; we comment on this in Appendix~\ref{app:sup}.

Our work asks the question of whether there exists causally ordered processes, which are not direct processes, but which can be considered as resources equivalent to the switch. Indeed, any advantages that the switch might have in Shannon theory would disappear if we allowed ourselves to compare it against arbitrary causally ordered processes such as 
\begin{equation}
\label{eq:circuit_swap}
\Qcircuit @C=0.5em @R=0.5em @!R {
&\qswap & \qw  &  \qw &\qw \\
& \qswap \qwx& \gate{\M_A} & \gate{\M_B}&\qw & 
},
\end{equation}
where the two-qubit gate is a swap. On the one hand, one might rightfully object that the causally order process~\eqref{eq:circuit_swap} lies outside the set of permitted processes since the sender makes use of a swap operation to directly transfer the information on a side-channel which is not affected by the action of the two noisy channels. On the other hand, it is also important to keep a relatively large class of causally ordered processes against which the process under consideration can be compared; otherwise any advantage would be empty of practical significance.

Are the above-mentioned communication advantages offered by the quantum switch a mere consequence of the presence 
of a side-channel similar to Eq.~\eqref{eq:circuit_swap}, or can they be meaningfully attributed to indefinite causality? We attempt to address this question by defining a class of processes that includes the quantum switch, as well as conceptually similar causally ordered processes, but excludes the process of Eq.~\eqref{eq:circuit_swap}. We will formalise our discussion using the process matrix formalism, but equivalent definitions could be provided in other formalisms, for example with quantum combs~\cite{Chiribella2009} in the causally ordered case, or using the Kraus representations that were favoured in Refs.~\cite{Ebler2017, Salek2018,Chiribella2018}.

\section{Process matrices} The process matrix formalism is a general framework for quantum mechanics that dispenses with the need to preassume a causal order between events. We quickly review the formalism here, and refer to the original references for a more complete introduction. A process matrix can be defined as a supermap~\cite{Chiribella2008} acting on local quantum channels~\cite{Araujo2017}. To each local laboratory (or party) is associated an input Hilbert space $\HH_{A_I}$ and an output Hilbert space $\HH_{A_O}$ (here $A$ is a label for the laboratory of Alice), with finite dimensions $d_{A_I}$ and $d_{A_O}$ respectively. A quantum channel $\M_A : \LL(\HH_{A_I}) \to \LL (\HH_{A_O})$ is a completely positive trace-preserving (CPTP) map, where $\LL(\HH)$ denotes the space of linear operators on $\HH$. It is convenient to define the Choi-Jamio\l kowski (CJ) isomorphism~\cite{Choi1975}, which associates to every quantum channel $\M$, a corresponding quantum state
\begin{equation}
M^{A_I A_O} := \sum_{i,j =1}^{d_{A_I}} |i \rangle \langle j|^{A_I} \otimes \M(|i \rangle \langle j|)^{A_O}
\end{equation}
It will also be useful to define a ``pure'' version of the CJ isomorphism: if $K : \HH_{A_I} \to \HH_{A_O}$ is a linear map, then
$
|K \rrangle^{A_I A_O} := \sum_{i=1}^{d_{A_I}} |i \rangle^{i} \otimes (K |i \rangle)^{A_O}.
$

We define here process matrices in the bipartite case (Alice and Bob), its generalisation to more parties is straightforward. Let $\HH_P, \HH_F$ be Hilbert spaces corresponding to the ``past'' and ``future'', respectively. A process is a bilinear \textit{supermap} that sends any pair of local quantum channels $\M_A , \M_B$, to a quantum channel $\mathcal{G}(\M_A, \M_B): \LL(\HH_P) \to \LL(\HH_F)$. Mathematically, a process matrix is an operator $W \in \LL( \HH_P \otimes \HH_{A_I} \otimes \HH_{A_O} \otimes \HH_{B_I} \otimes \HH_{B_O} \otimes \HH_F)$, such that
\begin{equation}
G^{PF} := \tr_{A_I A_O B_I B_O} \left(W^{T_{A_I A_O B_I B_O}} \cdot M^{A_I A_O} \otimes M^{B_I B_O} \right),
\label{eq:G_PF}
\end{equation}
is (the Choi operator of) a CPTP map from $\HH_P$ to $\HH_{F}$ whenever $M^{A_I A_O}$ and $M^{B_I B_O}$ represent CPTP maps. In Eq.~\eqref{eq:G_PF}, $T_{A_I A_O B_I B_O}$ designates the transposition on the spaces $\HH_{A_I}, \HH_{A_O}, \HH_{B_I}, \HH_{B_O}$ Hilbert spaces, while $\HH_P$ and $\HH_F$ are left untouched. A pure process~\cite{Araujo2017} is a process such that $G_{AB}$ is a unitary channel whenever $M^{A_I A_O}$ and $M^{B_I B_O}$ are unitary channels. It can be shown that all pure processes are rank one projectors $W = |w \rangle \langle w|$~\cite{Araujo2017}.

In what follows we restrict our attention to cases where $d_{A_I} = d_{A_O} = d_{B_I} = d_{B_O} =: d$. We call $W$ a \textit{direct pure process} (in opposition to a process with side channels), if $d_F = d_P = d$, and the induced channel $\mathcal{G}(\M_A,\M_B)$ is
\begin{equation}
\label{eq:direct_pure_AB}
\Qcircuit @C=1em @R=1em @!R {
&\gate{T}& \gate{\M_A} &\gate{U} & \gate{\M_B}& \gate{V}&\qw & 
}
\end{equation} 
or
\begin{equation}
\label{eq:direct_pure_BA}
\Qcircuit @C=1em @R=1em @!R {
&\gate{T}& \gate{\M_B} &\gate{U} & \gate{\M_A}& \gate{V}&\qw & 
},
\end{equation} 
for some fixed unitaries $T, U, V$. Such processes can be represented as rank one projectors: $W = |w \rangle \langle w|$. We will later consider $\M_A, \M_B$ noisy channels (e.g. depolarising channels). 
\begin{definition}
\label{def:SDPP}
\textit{Superpositions of direct pure processes (SDPP)} -- Let $\{|w_i \rangle^{P A_I A_O B_I B_O F}\}_{i=1}^{N}$ be a collection of pure direct processes. Then the (equal) superposition of these processes is defined by
$
|w \rangle = \frac{1}{\sqrt{N}} \sum_i |i \rangle^{C} |w_i \rangle^{P A_I A_O B_I B_O F},
$
which is a pure process with future Hilbert space $\HH_C \otimes \HH_F$. We call $\HH_C$ the control Hilbert space, and $\HH_F$ the target Hilbert space.
\end{definition}

It is clear that the quantum switch
\begin{align}
|w\rangle_{switch} = \frac{1}{\sqrt{2}} \bigg(&|0 \rangle^C |\id \rrangle^{P A_I}|\id \rrangle^{A_O B_I} |\id \rrangle^{B_O F} \nonumber \\
&+ |1\rangle^C |\id \rrangle^{P B_I}|\id \rrangle^{B_O A_I} |\id \rrangle^{A_O F} \bigg)
\end{align}
is a SDPP. However, it is also possible to take the superposition of two direct pure processes $|w_1\rangle$ and $|w_2\rangle$ with the \textit{same causal order}, for example with $A$ before $B$ in both cases. It seems that any reasonable resource theory that contains the quantum switch -- a superposition of direct pure processes with different causal orders -- should also allow superpositions of direct pure processes with the same causal order. If some communication advantage is not a generic property of SDPPs, but only of those processes which are causally non-separable~\cite{Araujo2015, Oreshkov2015, Branciard2016, Wechs2018}, this would yield credence to the claim that the advantage is due to indefinite causality. We note that all SDPPs can be implemented using interferometric setups very similar to those used for implementing the switch in Refs.~\cite{Procopio2015, Rubino2017, Goswami2018_switch, Wei2018}; we will return to this point in the Discussion.

We now proceed to show that there are causally ordered SDPPs that match (or outperform) the quantum switch for all tasks of Refs.~\cite{Ebler2017,Salek2018,Chiribella2018}.

\section{Classical and quantum communication} 
In Ref.~\cite{Ebler2017}, the authors consider two completely depolarising channels: $\M_A(\rho) = \M_B(\rho) = \tr(\rho) \frac{\id}{d}$. They find that placing these channels inside the switch allows for a non-zero classical channel capacity from $\HH_P$ to $\HH_C \otimes \HH_F$; if the target system has dimension two, they lower bound the channel capacity by $0.049$. Now consider the following two direct pure processes $|w_0 \rangle = |\id \rrangle^{P A_I} |\id \rrangle^{A_O B_I} |\id\rrangle^{B_O F}$ and $
|w_1 \rangle =  |X \rrangle^{P A_I} |\id \rrangle^{A_O B_I} |\id\rrangle^{B_O F}$, where $X$ is the Pauli-X matrix, as well as their corresponding superposition $|w\rangle = \frac{1}{\sqrt{2}}\left( |0\rangle^{C} |w_0\rangle + |1\rangle^C |w_1 \rangle \right)$. This process can be represented as the following circuit
\begin{equation}
\label{eq:circuit_cnot}
\Qcircuit @C=0.5em @R=0.5em @!R {
\lstick{\ket{+}} &\ctrl{1} & \qw  &  \qw &\qw \\
\lstick{} & \targ & \gate{\M_A} & \gate{\M_B}&\qw & 
},
\end{equation}
where the two-qubit gate is a controlled-not gate from the control to the target qubit, and $|+ \rangle = \frac{1}{\sqrt{2}} (|0 \rangle + |1 \rangle )$. This circuit allows for perfect communication of one classical bit: when the initial target state is $|+ \rangle$, the final state of the control is $|+\rangle$, while if the initial target state is $|-\rangle$, the final control state is $|-\rangle$. This is a manifestation of \textit{phase kick-back}, which has been identified as a unifying feature common to many quantum algorithms~\cite{Cleve1997}. Thus, we have found a causally ordered SDPP that outperforms the switch for the task of Ref.~\cite{Ebler2017}. We further note that a one-party SDPP would have been sufficient to obtain this effect (since removing $\M_B$ in circuit~\eqref{eq:circuit_cnot} does not change the final state), but we are sticking to the bipartite case to allow for direct comparison with the quantum switch.

In Ref.~\cite{Chiribella2018}, quantum communication through the channels $\M_A(\rho) = \M_B(\rho) = \frac{1}{2}\left( X\rho X + Y \rho Y\right)$, where $X$ and $Y$ are Pauli matrices, is considered. It is shown that inserting these channels in the quantum switch allows one to perfectly send one qubit from $\HH_P$ to $\HH_C \otimes \HH_F$. We now show that the process of Eq.~\eqref{eq:circuit_cnot} also allows perfect communication of one qubit, when the same channels are inserted in it.

First note that the composition of the two channels yields $\M_B (\M_A(\rho)) = \frac{1}{2} \left(\rho + Z \rho Z \right)$ i.e. a probability $\frac{1}{2}$ phase-flip channel. The process~\eqref{eq:circuit_cnot} plays a role analogous to that of a quantum error correcting code. The information about $|\psi\rangle$ can be recovered as follows: first measure the target qubit in the computational basis. If the outcome is $|0\rangle$, then it can be checked that the state of the control qubit will be $|\psi \rangle$. If we get the outcome $|1\rangle$, the state of the control qubit will be $X |\psi\rangle$, which can be corrected by applying a Pauli-X gate.

Finally, in Ref.~\cite{Salek2018} the authors consider the two channels $M_A(\rho) = (1- p)\rho + p X \rho X$ and $M_B(\rho) = (1 - q) \rho + q Z \rho Z$, and show that the quantum switch allows for a violation of the \textit{bottleneck inequality}, an inequality asserting that the quantum channel capacity of the concatenation of two channels $\M_A \circ \M_B$ is upper-bounded by the smallest of the quantum channel capacities of the two individual channels $\M_A$ and $\M_B$~\cite{Salek2018}. Consider the following causally ordered SDPP
\begin{equation}
\label{eq:circuit_cnot_mod}
\Qcircuit @C=0.5em @R=0.5em @!R {
\lstick{\ket{+}} &\ctrl{1} & \qw  & \qw &  \qw &\qw & \qw\\
\lstick{} & \targ & \gate{\M_B} & \gate{H} & \gate{\M_A}&\gate{H} & \qw  
}.
\end{equation}
The composition of $\M_B$, and $\M_A$ conjugated by Hadamard gates is a phase-flip channel. Therefore, the previous arguments establish that this process can be used to perfectly send one qubit of quantum information, thus violating the bottleneck inequality.

In Appendix~\ref{app:shor}, we present a causally ordered SDPP derived from the Shor quantum error correcting code~\cite{Shor1995, NielsenChuang}, that allows for perfect transmission of one qubit of information for all noisy channels $\M_A, \M_B$. This shows that taking the set of SDPPs as a resource (if one puts no limit on the number of control qubits) trivialises the problem of enhancing quantum and classical capacity.

\section{Discussion}
We now address the question of whether there are valid reasons to devise a resource theory that regards the quantum switch as an allowed process, while discarding general SDPPs. Indeed, one might be tempted to regard the circuit implementation of SDPPs, for example that in Eq.~\eqref{eq:circuit_cnot}, as an illegitimate resource for communication tasks, since the controlled-not gate allows some information to bypass the noisy channel (in other words, there appears to be signalling from the target qubit to the control qubit \textit{before the noisy channels are even applied}). To answer this question, we discuss here an interferometric implementation of the same process, for which such an objection does not apply, and that better emphasises its similarities with the quantum switch.

\begin{figure}[t]
\includegraphics[width=8cm]{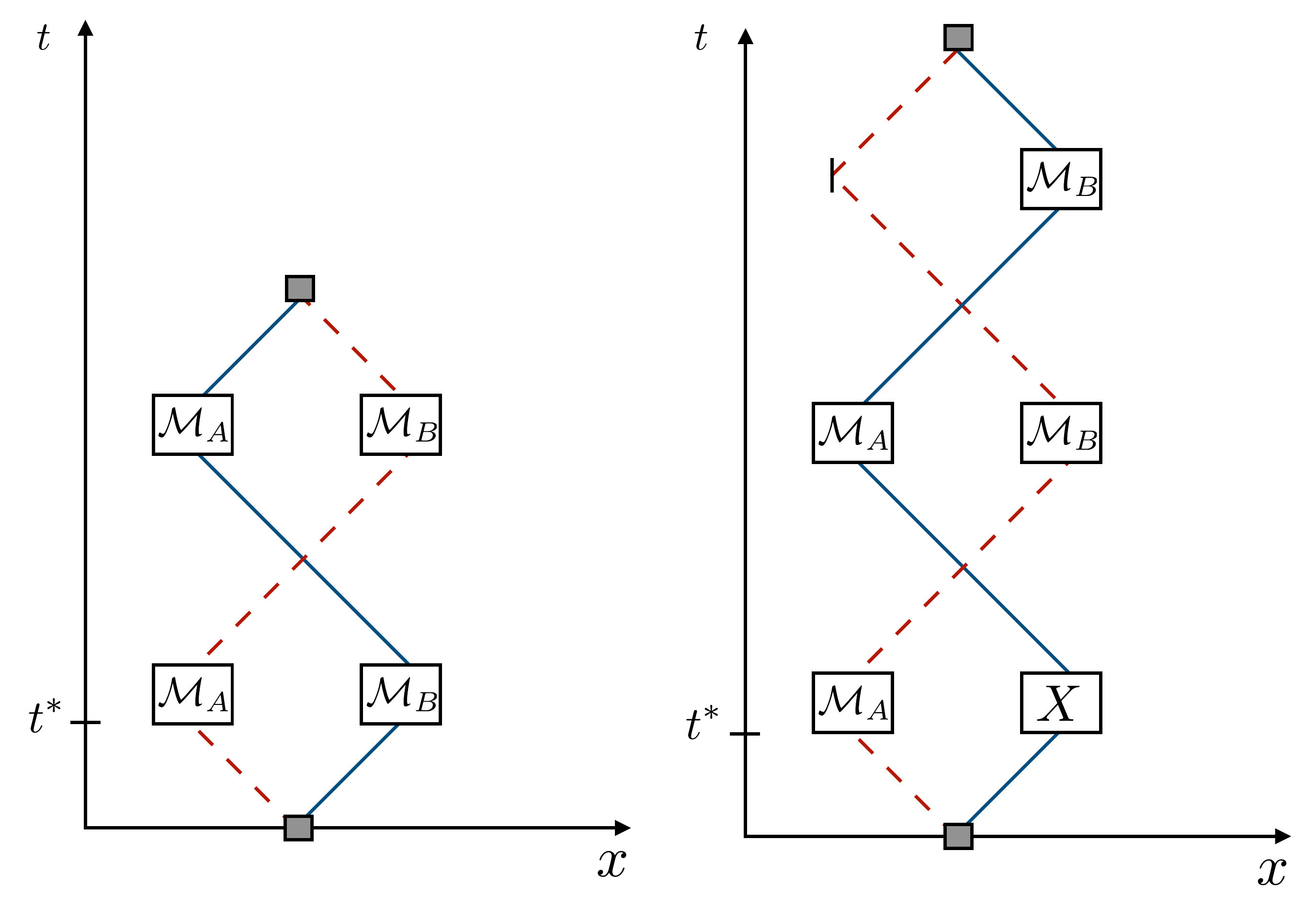}
\caption{\textbf{Comparison of the spacetime diagrams corresponding to implementations of the quantum switch (left), and the SDPP of Eq.~\eqref{eq:circuit_cnot} (right).} The small gray squares are beam splitters, and the dashed red and solid blue lines represent the two paths that are taken in an equal superposition by the photon. The noisy channels $\M_A, \M_B$, and the Pauli-X unitary act on the polarisation degree of freedom of the photon. There exists inertial reference frames in which the gate $X$ occurs at a later (or earlier) time than the first application of $\M_A$, which makes it impossible to assign a temporal order between the application of $\M_A$ on one side of the superposition and that of $X$ on the other side. Whether the action of X is part of the initial encoding or of the propagation channel depends on the reference frame.}
\centering
\label{fig:comparison}
\end{figure}

Figure~\ref{fig:comparison} compares the spacetime diagrams of an implementation of the quantum switch and of the causally ordered SDPP of Eq.~\eqref{eq:circuit_cnot}. A qubit is initially encoded in the polarisation degrees of freedom of a photon, which travels in an equal superposition of the two paths (blue and red). At the end, a localised observer can simultaneously access both modes and perform a measurement to recover information about the initial state of the polarisation qubit. The state of the photon at time $t^*$ (just before any noisy channel gets applied) is exactly the same in both implementations. Also note that both the target qubit and the control qubit are delocalised, and that the channel $\M_A$ at time $t^*$ only acts on the subspace spanned by $a^\dagger_{H} |\Omega\rangle$ and $a^\dagger_{V} |\Omega\rangle$, where $a^{\dagger}_{H/V}$ are the horizontally (resp., vertically) polarised modes corresponding to the path drawn in red, and $|\Omega \rangle$ is the vacuum state. A more detailed analysis of the nature of the systems which are acted upon by the channels in the switch was carried out in Ref.~\cite{Oreshkov2018}.

The recent work of Chiribella and Kristj\'{a}nsson~\cite{Chiribella2018_2nd} builds upon the previous studies~\cite{Ebler2017,Salek2018, Chiribella2018}. In this work the authors put forward a \textit{second-quantised Shannon theory}, in which both the internal and external degrees of freedom of the information carriers are quantised. They do not treat all degrees of freedom equally, but rather require a `\textit{clear-cut separation between the role of the internal and external degrees of freedom}'~\cite{Chiribella2018_2nd}, and in particular restrict the way that information can be distributed over internal and external degrees of freedom. While this restriction might serve the purpose of selecting the quantum switch out of all other possible SDPPs, it could be overly limiting if the goal is to understand the information-theoretical consequences of the fact that external degrees of freedom are also quantised.

Chiribella and Kristj\'{a}nsson thus set out to rule out processes such as that of Eq.~\eqref{eq:circuit_cnot}, which although (in our opinion) natural in the context of second-quantised information processing, trivialise the advantages claimed for the switch in Refs.~\cite{Ebler2017,Salek2018, Chiribella2018}. They do so by first defining an encoding channel $\mathcal{E}: \LL(\HH_Q) \to \LL(\HH_M \otimes \HH_P)$, of an abstract quantum system $\HH_Q$ into ``path'' $\HH_P$ and ``message'' $\HH_M$ degrees of freedom, and require it to satisfy a non-signalling condition
\begin{equation}
\label{eq:non_sig}
\tr_M (\mathcal{E}(\rho)) = \tr_M (\mathcal{E}(\sigma)), \forall \rho, \sigma \in \LL(\HH_Q).
\end{equation} 
This non-signalling requirement does not unambiguously achieve the goal of ruling out the undesired processes: consider once again the process at the right of Figure~\ref{fig:comparison}. It is clear from this space-time diagram that the unitary $X$ can equivalently be placed either before, simultaneously, or after the first application of $\M_A$ (and similarly for the first application of $\M_B$). This already shows an ambiguity regarding whether $X$ should be considered as part of the "encoding" or not. If we -- as we do here -- define the encoding to happen at time $t=0$ in the spacetime diagram, then the encoding is the same as for the switch: $\mathcal{E}(\rho) = \rho_M \otimes |+ \rangle \langle +|_P$, and it is non-signalling in the sense of Eq.~\eqref{eq:non_sig}. Just as it is ambiguous whether $X$ is part of the ``encoding'' operation, it is also unclear whether it should be considered, in the language of Ref.~\cite{Chiribella2018_2nd}, as a ``repeater operation'' (which are required to be non-signalling from the internal to the external degrees of freedom) or as a ``communication channel" (for which no such restriction applies). "Repeater operations" are defined as those parts of the quantum evolution that are not labelled as "communication channels", but it is not obvious what criteria determines which of the two labels applies for a given operation. 

Moreover, it is clear that both the switch and the SDPP in Fig.~\ref{fig:comparison} require two spatial modes to be implemented. In neither case is it possible for a single spatial mode to be used to transmit information, since no information is extractable from a single mode after all the noisy channels have been applied. The only way to send information through either of these two processes is to prepare a quantum superposition of the two spatial modes, and to recombine them after the noisy channels in order to read out the information. This suggests that there is no fundamental difference, regarding their status as a resource for Shannon theory, between the quantum switch and causally ordered SDPPs, and therefore, that the activation of channel capacity cannot be unambiguously attributed to indefinite causal order. Judging from Figure~\ref{fig:comparison}, a possible interpretation is that the counterintuitive channel capacities allowed by SDPPs are rather a consequence of the ability to delocalise noisy channels in space-time. This feature allows establishing correlations between the noise in the different arms of the interferometer, which can be useful in protecting against noise.

Finally, one might say that the arguments of this section only apply to a particular implementation, and that they do not apply to the quantum switch, when seen as an abstract primitive. Indeed our arguments have shown that the notion of "encoding" depends on the implementation, and it is not a property of an abstract process matrix. The abstract way to know whether a process contains a ``side-channel'' is to look at the reduced process~\cite{Araujo2015}, where Alice and Bob are traced out 
\begin{align}
W_{r} &=\frac{1}{4} \tr_{A_I A_O B_I B_O} |w \rangle \langle w|_{switch} \nonumber \\
& = \frac{1}{4}\id^{P C F} + \frac{1}{8} \left(|0 \rangle \langle 1| + |1 \rangle \langle 0| \right)^{C} \otimes |\id \rrangle \llangle \id|^{PF}.
\end{align}
We see that in the case of the quantum switch, the reduced process $W_{r}$ is a quantum channel which allows for direct communication from $\HH_P$ to $\HH_C \otimes \HH_F$, no matter what noisy operations are being applied at $A$ and $B$. As means of comparison, the reduced process for the SDPP of Eq.~\eqref{eq:circuit_cnot} is
\begin{equation}
W_r' = \frac{1}{4} \left(\id^{PCF} + X^P \otimes X^C \otimes \id^F \right).
\end{equation}
Thus if one accepts this abstract definition of a side-channel, both the quantum switch and the process of Eq.~\eqref{eq:circuit_cnot} contain a "side-channel" that allows direct signalling from $P$ to $F$. We leave open the question of whether the two side-channels should be considered equivalent, or whether they can be differentiated in a physically meaningful way.

\section{Conclusion}Refs~\cite{Ebler2017,Salek2018,Chiribella2018} have shown that the quantum switch allows for certain enhancements of classical and quantum channel capacity, when compared against a restricted class of causally ordered processes, and have coined the term \textit{causal activation} for this phenomenon. In this paper, we have examined whether these advantages can be univocally attributed to indefinite causality, or whether there are causally ordered processes that offer the same advantages and can be considered as equivalent resources. We have found insufficient reasons to reject any process belonging to a class that we named \textit{superpositions of direct pure processes}. Some of the processes belonging to that class are causally ordered, but can match  -- or outperform --  the quantum switch at the proposed tasks. We have also shown that the Shor quantum error correcting code can be used to find an SDPP that protects against arbitrary Pauli errors on the target qubit. Thus, we suggest that `\textit{controlled activation}' might be a more apt name for the above-mentioned phenomenon.

Our findings bear similarities with those of Ref.~\cite{Abbott2018}. In that work, an enhancement of channel capacity through quantum-controlled noisy channels is also found to be possible without indefinite causal order. However, the controlled-noise operations of Ref.~\cite{Abbott2018} are not process matrices, and thus their results depend on the specific implementation of the noisy channels. Our results however, as those of Refs.~\cite{Ebler2017, Salek2018, Chiribella2018}, are about process matrices and are independent of the way in which the noise is implemented. We elaborate on these comments in Appendix~\ref{app:sup}.

Let us conclude with an attempt at interpreting the difference between the previously discussed tasks, and other information processing tasks for which only an indefinite causal order yields an advantage~\cite{Chiribella2012, Araujo2014_NJP,  Araujo2014_PRL, Guerin2016}. A possible explanation for why there are causally ordered SDPPs that display a channel capacity activation is that in the protocols considered here the same noisy channels get applied in all runs of the protocol -- they are not subject to interventions by any parties. Instead, all previously known advantages of the quantum switch over causally ordered processes rely crucially on the possibility to controllably change the local operations across different runs~\cite{Chiribella2012, Araujo2014_NJP,  Araujo2014_PRL, Guerin2016}. This is a significant difference, as the operational definition of causality is formulated in terms of interventions.

\section*{Acknowledgements}
We thank Alastair Abbott, Mateus Ara\'ujo, Cyril Branciard, Rafael Chaves, Giulio Chiribella, Daniel Ebler, Ognyan Oreshkov, Lee Rozema and Sina Salek for helpful discussions. We acknowledge the support of the Austrian  Science  Fund  (FWF)  through  the  Doctoral  Programme  CoQuS, the uni:docs fellowship programme at the University of Vienna, and  the  projects  I-2526-N27  and  I-2906. This publication was made possible through the support of a grant from the John Templeton Foundation.  The opinions expressed in this publication are those of the authors and do not necessarily reflect the views of the John Templeton Foundation.

\appendix

\section{Superpositions of paths}
\label{app:sup}
In Ref.~\cite{Abbott2018}, it was already pointed out that the advantages of the switch for classical communication and for violating the bottleneck inequality can be also obtained through ``superpositions of paths''. The type of scenario that they considered is the following: let $\{K_i \}$ and $\{L_j \}$ be Kraus operators of the noisy channels $\M_A$ and $\M_B$, and consider the channel obtained from the isometry
\begin{align}
|\psi \rangle \mapsto &\frac{1}{\sqrt{2}} |0 \rangle^{C} \sum_i K_i |\psi \rangle^{T} |i \rangle^{E_0} |\epsilon_1\rangle^{E_1} \nonumber \\
&+ \frac{1}{\sqrt{2}} |1 \rangle^{C} \sum_i L_j |\psi \rangle^{T} |\epsilon_0 \rangle^{E_0} |j\rangle^{E_1},
\label{eq:isometry}
\end{align}
after tracing out the ancillary Hilbert spaces $E_0$ and $E_1$. In the above equation, $C$ designates the control degree of freedom, and $T$ the target. In Ref.~\cite{Chiribella2018}, it was shown that this type of ``superposition of paths'' does not  permit noiseless transmission of one qubit, while such a feat is possible using the quantum switch. The authors of Ref.~\cite{Chiribella2018} stated that their '\textit{findings highlight a fundamental difference between the type of interference arising from independent channels in a superposition of alternative paths and independent channels in a superposition of alternative orders}'. In what follows we attempt to clarify the previous statement, by comparing Eq.~\eqref{eq:isometry} with the SDPPs defined in the main text.

First, note that the isometry in Eq.~\eqref{eq:isometry} does not define a valid process matrix. Indeed,  the channel that one gets after tracing $E_0$ and $E_1$ depends on the particular Kraus definition of $\M_A, \M_B$~\cite{Oi2003, Abbott2018}. Therefore, that channel can not be obtained from a process matrix -- a bilinear map acting on the quantum channels $\M_A, \M_B$. In contrast, the SDPPs (which include the quantum switch) that we have been considering in this paper are indeed process matrices.

Second, it is important to make a distinction between \textit{independent channels} and \textit{correlated noise}. Both in Eq.~\eqref{eq:isometry} and for SDPPs, the channels $\M_A, \M_B$ are independent in the sense that they can be chosen independently. However, in Eq.~\ref{eq:isometry}, the noisy channels that get applied in each of the two paths are uncorrelated: the Kraus operators corresponding to $\M_A$ only appear in the path $|0\rangle^C$, and those corresponding to $\M_B$, only in $|1\rangle^C$. This is not the case for SDPPs: the noisy channels $\M_A, \M_B$ are correlated. Thus, what the findings of Ref.~\cite{Chiribella2018} highlight is rather a difference between interference arising from uncorrelated channels in a superposition of alternative paths, and interference arising from channels that are correlated via a superposition of direct pure processes.

\section{The Shor code as a SDPP}
\label{app:shor}
The Shor quantum error correction code~\cite{Shor1995, NielsenChuang} uses nine physical qubits to encode one logical qubit, in a way that protects it against arbitrary single qubit errors. 
\begin{equation}
\label{eq:circuit_Shor}
 \Qcircuit @C=0.5em @R=.5em {
   \lstick{|\psi\rangle} & \ctrl{3} & \ctrl{6} & \gate{H} & \ctrl{1} & \ctrl{2} & \multigate{8}{E} & \ctrl{1} & \ctrl{2} & \targ     & \gate{H} & \ctrl{3} & \ctrl{6} & \targ     & \rstick{\ket{\psi}} \qw\\
   \lstick{\ket{0}}      & \qw      & \qw      & \qw      & \targ    & \qw      & \ghost{E}        & \targ    & \qw      & \ctrl{-1} & \qw      & \qw      & \qw      & \qw       & \qw \\
   \lstick{\ket{0}}      & \qw      & \qw      & \qw      & \qw      & \targ    & \ghost{E}        & \qw      & \targ    & \ctrl{-2} & \qw      & \qw      & \qw      & \qw       & \qw \\
   \lstick{\ket{0}}      & \targ    & \qw      & \gate{H} & \ctrl{1} & \ctrl{2} & \ghost{E}        & \ctrl{1} & \ctrl{2} & \targ     & \gate{H} & \targ    & \qw      & \ctrl{-3} & \qw \\
   \lstick{\ket{0}}      & \qw      & \qw      & \qw      & \targ    & \qw      & \ghost{E}        & \targ    & \qw      & \ctrl{-1} & \qw      & \qw      & \qw      & \qw       & \qw \\
   \lstick{\ket{0}}      & \qw      & \qw      & \qw      & \qw      & \targ    & \ghost{E}        & \qw      & \targ    & \ctrl{-2} & \qw      & \qw      & \qw      & \qw       & \qw \\
   \lstick{\ket{0}}      & \qw      & \targ    & \gate{H} & \ctrl{1} & \ctrl{2} & \ghost{E}        & \ctrl{1} & \ctrl{2} & \targ     & \gate{H} & \qw      & \targ    & \ctrl{-6} & \qw \\
   \lstick{\ket{0}}      & \qw      & \qw      & \qw      & \targ    & \qw      & \ghost{E}        & \targ    & \qw      & \ctrl{-1} & \qw      & \qw      & \qw      & \qw       & \qw \\
   \lstick{\ket{0}}      & \qw      & \qw      & \qw      & \qw      & \targ    & \ghost{E}        & \qw      & \targ    & \ctrl{-2} & \qw      & \qw      & \qw      & \qw       & \qw
 }
\end{equation}
where $E$ is an arbitrary single-qubit error channel.
The Shor code can be interpreted as a SDPP, by using circuit identities to flip the direction of the controlled gates. Furthermore, if we are merely interested in correcting errors on the target qubit only (the top qubit in the circuit above), we can remove qubits number $5,6,8,9$. The following circuit is a simplified version of the Shor code (where we omit the part of circuit~\eqref{eq:circuit_Shor} that happens after the error channel), and it can be implemented as a uniform superposition of $2^4$ processes with the same causal order. Since we are interested in correcting errors on the target qubit only (the lowest qubit in circuit~\eqref{eq:circuit_shor_SDPP}), it turns out we can use only 5 physical qubits.
\begin{equation}
\label{eq:circuit_shor_SDPP}
\Qcircuit @C=0.5em @R=0.5em @!R {
\lstick{\ket{+}} &\qw & \qw   &\qw & \qw & \ctrl{4}  & \qw & \qw & \qw & \qw \\
\lstick{\ket{+}} &\qw & \qw   &\qw & \ctrl{3} & \qw  & \qw & \qw & \qw & \qw\\
\lstick{\ket{+}} &\qw & \qw   & \ctrl{2} & \qw & \qw & \qw & \qw & \qw & \qw\\
\lstick{\ket{+}} & \qw & \ctrl{1} & \qw   &\qw & \qw & \qw & \qw & \qw & \qw\\
\lstick{|\psi\rangle} & \gate{H} & \targ & \targ & \targ & \targ & \gate{H} &  \gate{\M_A} &  \gate{\M_B} & \qw  
}.
\end{equation}
This SDPP allows to perfectly transmit one qubit of quantum information, for all channels $\M_A, \M_B$. Thus, the SDPP of Eq.~\eqref{eq:circuit_shor_SDPP} generalises, and improves upon, the observations made in the main text. 

This example shows that SDPPs (if one puts no limit on the number of control qubits) can be used to perfectly send one qubit of information, essentially trivialising the problem of enhancing quantum and classical channel capacity if one were to take the set of all SDPPs as a resource.


\newpage

\providecommand{\href}[2]{#2}\begingroup\raggedright\endgroup

\end{document}